\documentstyle[preprint,aps,fixes,epsf]{revtex}
\begin{document}
\preprint{\hfill{\parbox{4.0cm}{UW/PT-96-07\ \\  SNUTP 96-045
\ \\hep-ph/9607464
}}}
\title{Cone Algorithm Jets in $e^+e^-$ Collisions}
\author{Junegone Chay\footnote{e-mail address: \tt
chay@kupt.korea.ac.kr}} 
\address{Department of Physics, Korea University, Seoul 136-701,
Korea}
\author{ Stephen D. Ellis\footnote{e-mail address:
\tt ellis@phys.washington.edu}}
\address{Department of Physics, FM-15, University of Washington, Seattle,
WA~98195} 
\date{\today}
\maketitle
\begin{abstract}
The structure of hadronic jets depends not only on the dynamics of QCD but 
also on the details of the jet finding algorithm and the physical process in 
which the jet is produced.  To study these effects in more detail
we calculate the jet cross section and the internal jet structure in
$e^+e^-$ annihilations and compare them to the results found in
hadronic collisions using the {\em same} jet 
definition, the cone algorithm. The different 
structures of the overall events in the two cases are evident in the
comparison. For a given cone size and jet energy, the 
distribution of energy inside the cone is more
concentrated near the center for jets from $e^+e^-$ collisions than 
for jets from hadronic collisions.  
\end{abstract}
\pacs{12.38.Bx, 12.38.Qk, 13.87.-a}
\narrowtext
\section{Introduction}
\label{intro}
Jets of narrowly collimated energetic hadrons are clearly seen in high
energy collisions. They are observed in $p\overline{p}$
collisions\cite{ppbar} (CERN, Fermilab), in deep inelastic $ep$
scattering\cite{ep} (HERA) and in $e^{+}e^{-}$
annihilations\cite{ee} (SLAC, LEP, DESY, KEK). It is important to know
how to analyze these jets quantitatively since they are essential not
only for understanding and testing the underlying strong interaction theory
but also in looking for new physics beyond the Standard Model such as
the Higgs boson.  The goal is to be able to employ the jets as surrogates 
for the underlying quarks and gluons to quantitatively characterize event
structures in much the same way that leptons are used.  Thus the main issue 
in studying jets is to reduce the various uncertainties, both
theoretical and experimental.

The theoretical uncertainties in studying jets come from various
sources. In the $p\overline{p}$ case  the largest uncertainty comes
from incomplete knowledge of the parton distribution functions,
especially the gluon distribution function at small $x$. The jet cross
section can vary by at least 10\% when different sets of distribution
functions are employed, although this situation is improving with time. 
Clearly this uncertainty is absent in
$e^{+}e^{-}$ collisions. Secondly, there is uncertainty associated
with the uncalculated higher-order corrections. This is illustrated by
the fact that the theoretical cross section exhibits a dependence on
the unphysical and arbitrary renormalization (factorization) scale,
$\mu$. While the experimental jet cross section is, of course,
independent of this scale, the residual dependence in the fixed-order
perturbative result is a remnant of the truncation of the perturbative
expansion. 

Calculations\cite{sellis,greco} have been performed using the
matrix elements at next-to-leading order\cite{kellis} in
$p\overline{p}$ collisions. When the next-to-leading order terms are
included, the dependence of the cross section on the renormalization
scale is markedly reduced, suggesting an uncertainty of order 10\%
due to the uncalculated higher-order corrections.
For $e^+e^-$ collisions the situtation is somewhat different.  At
lowest order ($\alpha_s^0$) there is no $\mu$ dependence.  Only at
order $\alpha_s^1$, as addressed here, does $\mu$ dependence appear.
At this order there can be no cancellation of the $\mu$ dependence with 
higher orders.
However, for the issue of the internal structure of jets, the real
focus of this study, the dependence on $\mu$ should be comparable in
the two cases, $p{\bar p}$ at order $\alpha_s^3$ and $e^+e^-$ at order
$\alpha_s^1$.  In both cases the structure is evaluated at lowest
non-trivial order.

Another important source of uncertainty arises from the use of
different theoretical and experimental jet definitions. We can obtain
an appreciation of this issue by considering the following qualitative
pictures for jet productions.  In $p\overline{p}$
collisions, two of the the partons in the incoming hadrons undergo a hard
scattering producing final-state partons with large momenta
transverse to the beam direction. The scattered partons can radiate
further partons both after the hard scattering, final-state radiation,
and before the scattering, initial-state radiation. The hard
scattering process is also immersed in the background that arises from
the interactions of the spectator partons. This underlying event is
not part of the hard scattering but does contribute to the overall
event. The partons from all of these sources then participate in the
less-well-understood process of fragmentation into hadrons and can
participate in the formation of a jet. 

In $e^{+}e^{-}$ collisions,
electrons and positrons annihilate to produce initially a small number
of energetic partons. These partons then radiate more partons that all
fragment into hadrons that can be associated with jets. Since
there are no incoming partons, the initial-state radiation and the 
background due to the spectator partons are absent. In both kinds of 
collisions the issues of color and energy-momentum conservation ensure
that the fragmentation process into hadrons is a collective one with
large numbers of partons acting coherently. Thus there can be no
unique identification of a set of hadrons or a jet with a single
scattered parton. Since there is no unique {\it a priori} jet
definition for combining the final particles to form a jet, there is
an arbitrariness in the choice of jet algorithm and different choices
produce different results for the same process.

The dependence of jet cross sections on the jet definition is an issue
both for comparing different experiments and for comparing experiment
to theory. Precise comparisons are possible only if the detailed
dependence on the jet definition exhibited in the data is reproduced
by the theory. This is a limitation of the Born level calculation,
which has only one parton per jet, thus exhibiting no structure to the
jet at all. In Monte Carlo simulations the finite size of a jet arises
from the subsequent showering of this parton and the nonperturbative
fragmentation into hadrons. Both of these features contribute to the
theoretical uncertainty. The inclusion of higher-order corrections
reduces this uncertainty in the jet definition dependence because we
can see nontrivial jet structure in the perturbative calculation.
However, there remain many questions about the contribution of the
rest of the event to the jet and to the uncertainty in the cross
section. The interplay of the jet definition with the initial-state
radiation and the underlying event can be studied by comparing two jet 
samples, one from $p\overline{p}$ collisions and the other from
$e^{+}e^{-}$ annihilations, where these two contributions are
different. The goal is to generate experimental jet samples from
$e^{+}e^{-}$ events that can be directly compared with those from
hadronic collisions.  In this paper we compare the theoretical results 
for jets in $e^{+}e^{-}$ and $p\overline{p}$ collisions 
using the same jet definition in each case.

In Section II, we discuss kinematic differences in $e^{+}e^{-}$
collisions and $p{\overline p}$ collisions. In Section III, we define
a jet in $e^{+}e^{-}$ and hadronic collisions. In Section IV, the
characteristics of the jet cross section in $e^{+}e^{-}$ collisions
are discussed. We compare the transverse energy distribution of
the jets in $e^{+}e^{-}$ and hadronic collisions in Section V. In the
final section, we summarize the features of jets in both cases. 

\section{Kinematic Differences}

Let us first consider the kinematic differences in $e^{+}e^{-}$
collisions and $p{\overline p}$ collisions. In the $e^{+}e^{-}$ case,
the center-of-mass energy of the partons participating in the hard
scattering, $\sqrt{\widehat s}$, is fixed and equal to the total
energy $\sqrt{\widehat{s}}=\sqrt{s}=Q$. Also, since the electrons and
positrons have equal and essentially opposite momentum, the laboratory
frame and the center-of-mass frame coincide. As a result the event
structure is essentially spherical with respect to the interaction point
and the detector geometry tends to exhibit the same symmetry.
Generally very simple jet definitions have been employed in
$e^{+}e^{-}$ experiments. For two-jet events, a jet is simply a
hemisphere. For multijet events with a small number of jets simple
invariant mass cuts have been used to define jets.

The $p\overline{p}$ case is more complex. The center-of-mass energy of
the hard parton scattering is given by ${\widehat s}=x_1x_2s$ where
$\sqrt{s}$ is the center-of-mass energy of the beam particles and
$x_1$ and $x_2$ are the fractions of the longitudinal momenta of 
the incoming hadrons carried by the 
scattering partons.  Although the 
incoming hadrons have equal and opposite momentum, the scattered
partons in general do not and the center-of-mass frame of the parton
scattering is boosted along the beam direction with respect to the
laboratory frame. Thus the relevant phase space is effectively
cylindrical in the laboratory and the detectors for hadronic collider
experiments are designed to match this symmetry. Likewise the natural
variable is the transverse energy, $E_T$, which is the component of
the energy perpendicular to the beam axis and which is invariant under
boosts along the beam direction. Since this characteristic is
intrinsic to $p\overline{p}$ collisions, we have chosen to employ the
jet definition of $p\overline{p}$ collisions both in $e^+e^-$ and
$p\overline{p}$ collisions to compare the jet samples in both cases. 

An important kinematic difference between the $e^{+}e^{-}$ and the 
$p\overline{p}$ cases arises from the fact, noted above, that in the
$e^{+}e^{-}$ case one performs experiments at fixed energy for the 
hard scattering,
$\sqrt{s}=\sqrt{\widehat{s}}$. Thus fixing the energy of a jet imposes
a strong constraint on the rest of the hard scattering event and thus
on the full final state. For example, if we fix the energy of the jet,
we know how much energy associated with the hard scattering is outside
the jet. However, this is not the case for hadronic collisions.
Although we fix the energy of the jet, we do not know precisely how
much energy associated with the hard scattering is outside the jet.
The energy of the hard scattering process is $\sqrt{{\hat s}}=$
$\sqrt{x_1x_2s}$ and depends on the specific $x_i$ values and thus on
the parton structure functions. In general, however, the distribution
functions are rapidly falling functions of the $x_i$ and
$\sqrt{\widehat{s}}$ will be only slightly larger, on average, than
necessary to generate the chosen jet. Another point is that, since we
have chosen to impose the variable $E_T$ on the $e^{+}e^{-}$ case, we
will have to integrate over the polar angle $\theta $, or in the
hadronic language the pseudorapidity $\eta =\ln \cot \theta /2$, in
order to make $E_T$ a free variable at lowest order. These differences in the
kinematics will be important for understanding some of the ``trivial''
differences between the jets observed in the two types of experiments.

\section{Jet Definition}

Since jet cross sections critically depend on the jet definitions
used, we can compare jet cross sections from different experiments
only if we use the same jet definition for both of the jet samples.
Thus we want to establish a standard jet definition. The point is not
to select an optimal jet definition, since that will depend on
specific applications, but to formulate a jet definition that
satisfies reasonable criteria and can be used by both experimentalists
and theorists to generate a sample of jets or jet cross sections for a
wide range of processes that can be meaningfully compared between
different groups. The relevant criteria are that the jet definition be
easy to implement in all experiments and theoretical calculations, and
yield reliable, finite results at any order in perturbation
theory\cite{snowmass}.

As discussed briefly above, the jet definition employed in practice in 
hadron collisions is characterized in terms of the transverse energy
$E_T= E\sin \theta $ measured inside a cone in $\eta $-$\phi $ space,
where $\eta =\ln \cot \theta /2$ is the pseudorapidity and $\phi $ is
the azimuthal angle around the beam direction. In terms of calorimeter
cells, $i$, inside a cone defined by 
\begin{equation}
\label{jetcone}\Delta R_i\equiv \sqrt{(\eta _i-\eta _J)^2+(\phi
_i-\phi _J)^2 }\leq R,
\end{equation}
we define the transverse energy of the jet, $E_T$, as 
\begin{equation}
\label{transe}E_T=\sum_{i\in {\rm cone}}E_{T,i}.
\end{equation}
The energy weighted direction of the jet is given by 
\begin{equation}
\label{etaj}\eta _J={\frac 1{{E_T}}}\sum_{i\in {\rm cone}}E_{T,i}
\eta_i 
\end{equation}
and 
\begin{equation}
\label{phij}\phi _J={\frac 1{{E_T}}}\sum_{i\in {\rm cone}}E_{T,i}
\phi_i. 
\end{equation}
This procedure implies some number of iterations of the jet defining
process in Eq.\ (\ref{jetcone}) until the quantities defined in
Eqs.\ (\ref{transe}), (\ref{etaj}) and (\ref{phij}) are stable with
the jet cone remaining fixed. We can also apply this jet definition at
the parton level to form a jet. A single isolated parton with
$(E_T,\eta_J, \phi _J)$ can be reconstructed as a jet. Or two partons
with $(E_{T,1}, \eta _1,\phi _1)$ and $(E_{T,2},\eta_2,\phi _2)$ may
be combined into a single jet. In that case the jet transverse energy
is $E_T=E_{T,1}+E_{T,2}$, and $\eta _J=(E_{T,1}\eta_1+E_{T,2}\eta
_2)/E_T$, $\phi _J=(E_{T,1}\phi _1+E_{T,2}\phi _2)/E_T.$ To determine 
if the two partons are to be combined into a jet, we see if the two
partons are in a cone of radius $R$ about the jet axis. The condition
that parton 1 fits into the cone is $(\eta _1-\eta _J)^2+(\phi _1-\phi
_J)^2< R^2$, or 
\begin{equation}
\label{onecone}{\frac{{E_{T,2}}}{{E_{T,1}+E_{T,2}}}}|{\bf \Omega }_1- 
{\bf \Omega }_2|<R,
\end{equation}
where we denote a two dimensional vector ${\bf \Omega }=
(\eta ,\phi)$. Similarly, the condition that parton 2 fits in the cone
is  
\begin{equation}
\label{twocone}{\frac{{E_{T,1}}}{{E_{T,1}+E_{T,2}}}}|{\bf \Omega }_1- 
{\bf \Omega }_2|<R.
\end{equation}
Thus the combined condition is 
\begin{equation}
\label{bothcone}|{\bf \Omega }_1-{\bf \Omega
}_2|<{\frac{{E_{T,1}+E_{T,2}}} {{{\rm max}(E_{T,1},E_{T,2})}}}R.
\end{equation}
If the two partons satisfy this condition, then we count one combined jet
as specified above, but not the two one-parton jets. We can clearly  
generalize this definition to include more partons at higher-orders. 

This is the jet definition used for $p\overline{p}$ collisions in
Ref.\cite{sellis}.  For the analysis described here, where we do not attempt to 
describe the experimental data, we 
neglect the subtleties of the jet merging problem and the
parameter $R_{sep}$ discussed in the last paper in 
Ref.\cite{sellis}.  A detailed comparison to data will be presented 
separately\cite{chayellis}.  
The precise definitions\cite{jetal} used in the actual experiments
are similar to the definition used here.  Here we apply this definition to 
calculating jet cross sections in
$e^{+}e^{-}$ events in order to compare to the $p\overline{p}$ case.
As argued earlier, it is necessary to use the same definition of jets
in comparing jets from different sources. We also define the kinematic
variables $E_T=E\sin \theta $, the transverse energy perpendicular to
the electron beam direction, $\eta $, the pseudorapidity, and  $\phi
$, the azimuthal angle around the electron beam direction, in direct
analogy to the $p\overline{p}$ case (even though it is less natural in
the $e^{+}e^{-}$ case).

\section{Jet Cross Section}

We first consider the single inclusive jet cross section 
$d\sigma/dE_T$. Due to 
the kinematic constraints characteristic of $e^+ e^-$ collisions, this cross 
section exhibits a remarkable level of structure.  As 
discussed above, we choose $E_T$ as a variable instead of $\theta$, 
the angle from the beam direction. The differential Born cross section  
$(d\sigma/dE_T)_B$ (normalized to 2 jets per event) is given by
\begin{equation}
\label{bornet}\bigl(\frac{d\sigma}{dE_T}\bigr)_B=48\pi \alpha^2 
\sum_q e_q^2  
\frac{E_T(1-2E_T^2/{\widehat s})}{{\widehat s}^2\sqrt{1-4E_T^2
/{\widehat s}}},
\end{equation}
where $e_q$ is the quark charge, while the differential Born cross  
section $(d\sigma/d\Omega)_B$ for $e^+e^-\rightarrow q{\overline q}$  
is well known as
\begin{equation} 
\label{bornomega}\bigl(\frac{d\sigma}{d\Omega}\bigr)_B  =
\frac{3\alpha^2 }{4{\widehat s}} \sum_q e_q^2 (1+\cos^2 \theta).
\end{equation}

At order $\alpha _s$ in $e^{+}e^{-}$ collisions, nontrivial jet
structure appears. We now include real gluon emission from a quark or an 
antiquark and virtual gluon corrections to quark-antiquark pair
production. Final-state partons are combined to form a jet with a
finite size according to the jet definition defined in the previous
section. In the case of a three-parton final state the 
third parton can either be 
inside the detected jet or part of the system recoiling from the detected jet, 
which is constrained by energy-momentum conservation. To
actually evaluate the cross section, we organize the calculation by
adding and subtracting simplified matrix elements 
that have the correct divergences.
The singular pieces are evaluated analytically and are explicitly  
canceled. The remaining finite integrals are evaluated numerically.  
At order $\alpha_s$, the three partons are $q$, ${\overline q}$  
and $g$ and we label them as parton 1, 2 and 3, summing over all
possible identifications with the three partons. We
organize the calculation in such a way that parton 1 is opposite to
the jet direction to balance momentum and conserve energy and parton 3 
has the smallest transverse energy ($E_{T,1},\ E_{T,2} > E_{T,3}$). 
There are three possibilities of forming a jet for the jet cone size 
$R < \pi /3$. Either parton 2 alone or parton 3 alone can form a jet,
or partons 2 and 3 together form a jet. We choose the renormalization
scale $\mu = E_T/2$ as suggested by the $p{\bar p}$
calculations\cite{sellis}.  

\begin{figure}[t]
\vskip -1.0in
\vbox
    {%
    \centerline
    { \epsfbox{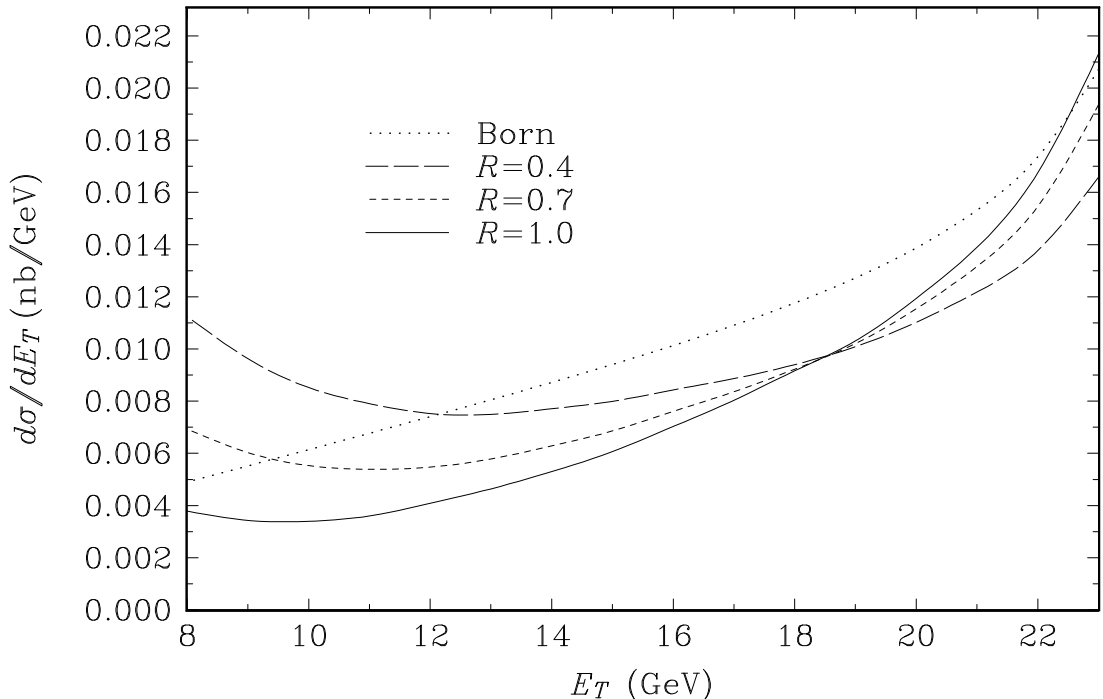} }
\vskip -0.25in
\caption{${d\sigma/dE_T}$ at order $\alpha_s$
with $\mu = E_T/2$ versus $E_T$ at 
$\protect\sqrt{{\widehat s}}=50$ GeV in
$e^+e^-$ collisions 
for $R=$ 0.4, 0.7 and 1.0 and the Born
cross section.}  
\label{figure1}
    }%
\end{figure}

Since we impose the jet algorithm, $d\sigma/dE_T$ depends on $R$ as 
well as $E_T$. Let us first consider $d\sigma/dE_T$ versus $E_T$
at fixed $R$.
Fig.\ \ref{figure1} shows $d\sigma/dE_T$ at $R=0.4$, 0.7 and 1.0 
respectively for $\sqrt{{\widehat s}}=50$ GeV along with the Born
cross section $(d\sigma/dE_T)_B$. This choice of energy is essentially 
arbitrary except that it is large ($\sqrt{{\widehat s}} \gg \Lambda_{QCD}$)
and serves to remind us that we have included only photon exchange 
and not $Z$ exchange.  We expect that jet production from $Z$  
exchange is very similar to the case of photon exchange \cite{brown}.  We
also note that, except for the scale $\Lambda_{QCD}$ in $\alpha_s$, this 
theoretical cross section is scale-free.  Results for other energies can be 
obtained (to a good approximation) by simply scaling the energy, keeping 
dimensionless ratios and angles fixed.  

\begin{figure}
\vskip -0.9in
\centerline
    { \epsfbox{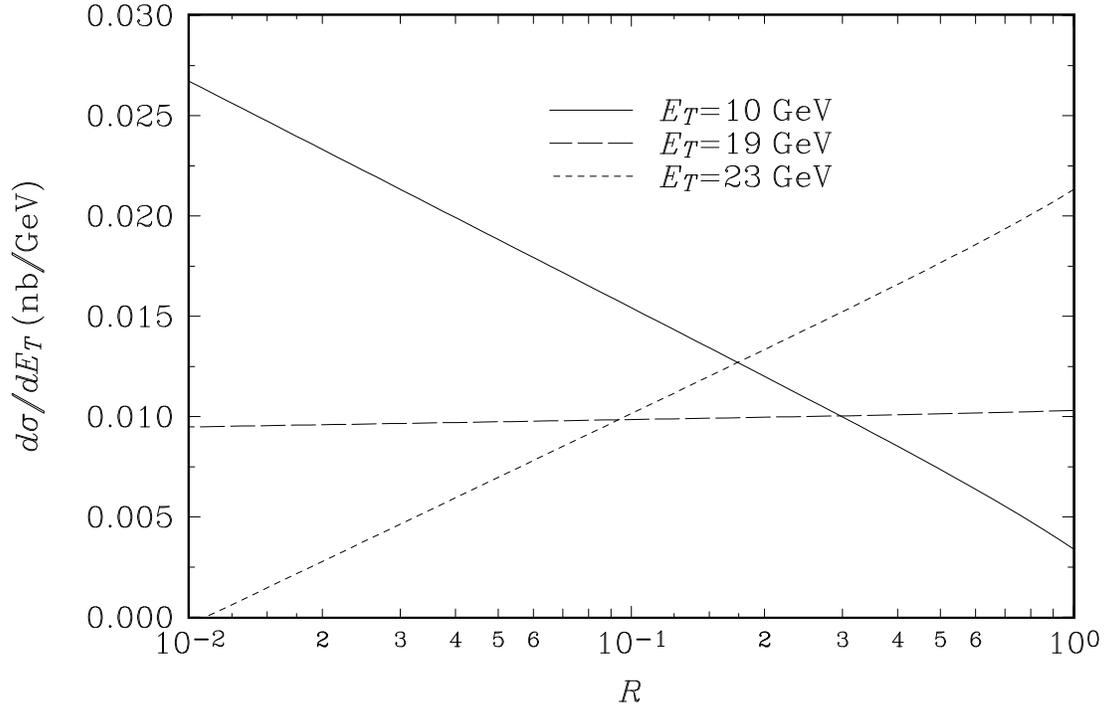} }
\vskip -0.25in
\caption{$d\sigma/dE_T$ at order $\alpha_s$ 
with $\mu = E_T/2$ versus $\ln R$ at 
$\protect\sqrt{{\widehat s}}=50$ GeV in
$e^+e^-$ collisions for $E_T=$10 GeV, 19 GeV 
and 23 GeV.} 
\label{figure2}
\end{figure}

As is already evident in 
Eq.\ (\ref{bornet}) the Born cross section diverges as
$E_T\rightarrow \sqrt{{\widehat s}}/2$ due to the Jacobian arising from
the change of variables from $\cos \theta$ to $E_T$. As we increase the
cone size, the cone tends to include more partons, thus more
transverse energy inside the jet. Therefore the cross section
increases for fixed $E_T$ close to $\sqrt{{\widehat s}}/2$ as we increase
$R$, while it decreases for fixed $E_T$ small compared to
$\sqrt{{\widehat s}}/2$ ({\it i.e.}, it becomes more difficult to keep
extra $E_T$ out of the cone).  As indicated in Fig.\ \ref{figure1} the 
transition between the two types
of behavior occurs for $E_T \simeq 19$ GeV or 
$x_T = 2E_T/\sqrt{\widehat s} \simeq 0.76$, where the cross section
is essentially independent of $R$.
This same feature is illustrated in Fig.\ \ref{figure2}
where the theoretical inclusive cross section for single jet
production at order $\alpha_s$ is plotted as a function of the cone
radius $R$ for $E_T=$10, 19 and 23 GeV respectively with
$\sqrt{\widehat{s}}=50$ GeV. While the Born result is independent of
$R$, the order-$\alpha _s$ cross section clearly depends on $R$, but
with the form of the dependence varying with $E_T$. The
slope with $R$ is positive for large $E_T$, negative for small $E_T$
and approximately vanishes for $E_T \simeq 19$ GeV or $x_T \simeq 0.76$.

\begin{figure}
\vskip -1.0in
\centerline
    { \epsfbox{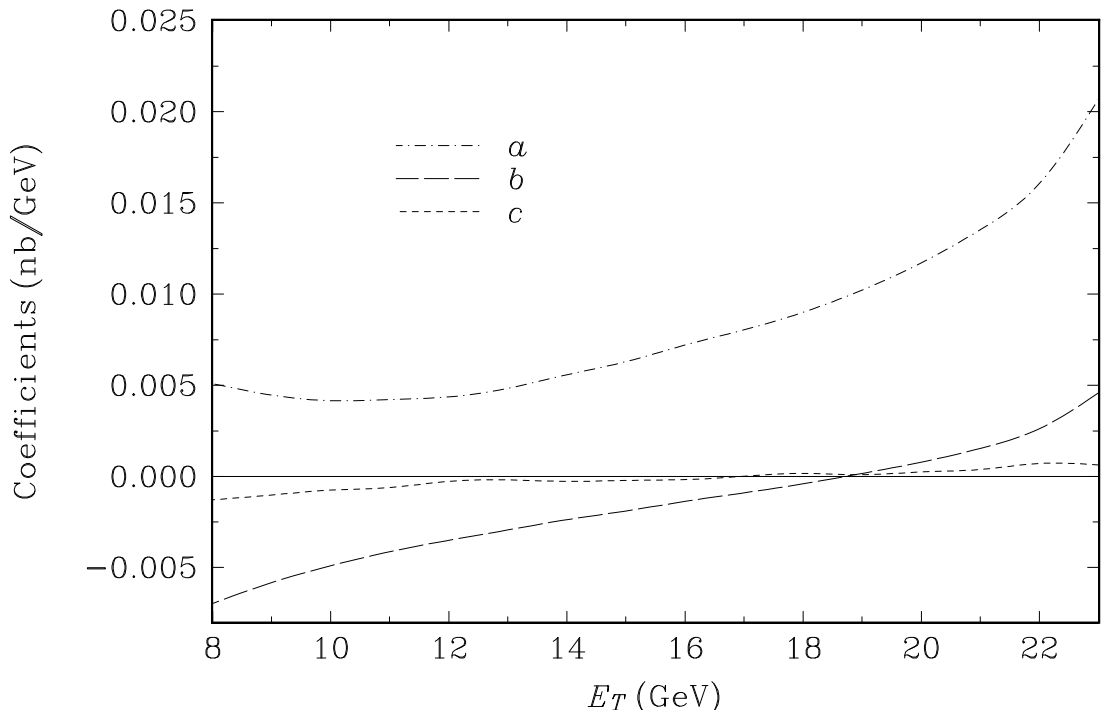} }
\vskip -0.25in
\caption{Coefficients $a$, 
$b$ and $c$ of Eq.\ (\protect\ref{rdepen}) 
for $e^+e^-$ collisions 
at $\protect\sqrt{{\widehat s}}=50$ GeV
as functions of $E_T$ with $\mu = E_T/2$.}
\label{figure3}
\end{figure}

The general form of this dependence is approximately characterized by
three parameters as 
\begin{equation}
\label{rdepen}{\frac{{d\sigma }}{{dE_T}}}\approx a+b\ln R+cR^2.
\end{equation}
The numerical values and $E_T$ dependence of the coefficients $a$, $b$
and $c$ are indicated in Fig.\ \ref{figure3} for $\sqrt{{\widehat
s}}=50$ GeV as in Figs.\ \ref{figure1} and \ref{figure2}.  We can
think of the parameter $a$ as describing the contribution of the 
2-parton final state, which is independent of $R$. Since this term is
dominated by the Born cross section, its $E_T$ dependence is easily
understood by comparing with Fig.\ \ref{figure1}. 

The form of the second term in Eq.\ (\ref{rdepen}) arises from the
collinear divergence present in the perturbation theory at order $\alpha_s$.
The sign of $b$ is negative for small $E_T$ and changes sign
as we increase $E_T$. This behavior of $b$ can be qualitatively
understood using the results for the Sterman-Weinberg
jet\cite{stwein}.  Sterman and Weinberg introduced a quantity $\sigma
(\theta ,\epsilon ,\delta )$, which is the cross section for
$e^{+}e^{-}\rightarrow q{\overline q}g$ where a fraction $(1-\epsilon
)$ of the total energy is emitted within two oppositely directed cones
of half-angle $\delta $, making an angle $\theta $ with the beam axis.
Both $\epsilon $ and $\delta $ are very small. The result to order
$\alpha _s$ is 
\begin{eqnarray}
\label{stweinberg}{\frac{{d\sigma }}{{d\Omega }}}(\theta ,\epsilon
,\delta )= \bigl({\frac{{d\sigma }}{{d\Omega }}}\bigr)_B
\bigl[1&-&{\frac{{4\alpha _s}}{{3\pi }}}\{(3+4\ln 2\epsilon )
\ln \delta \nonumber \\ 
&+&{\frac{{\pi ^2}}3}-{\frac 52}\}+O(\epsilon ,\delta )\bigr],
\end{eqnarray}
\noindent where $(d\sigma /d\Omega )_B$ is the Born differential cross
section given in Eq.\ (\ref{bornomega}). We can transform
Eq.\ (\ref{stweinberg}) to $d\sigma /dE_T$ using the relation
$E_T=E\sin \theta $. Note that the jet algorithm used by Sterman and
Weinberg is different from the jet algorithm we use here. However,
when $\delta $ and $R$ are small, $\delta $ is proportional to $R$. Then the
coefficient of $\ln R$ is the same as the coefficient of 
$\ln \delta$. 

For simplicity, let us consider the case in which $\theta $
approaches $\pi/2$ so that $E_T \rightarrow \sqrt{{\widehat s}}/2$. 
Then the relation between $E_T$ and $\epsilon$ becomes 
$2E_T=(1-\epsilon)\sqrt{{\widehat s}}\sin \theta \approx
(1-\epsilon )\sqrt{{\widehat s}}$.
When we replace $\epsilon $ in Eq.\ (\ref{stweinberg}) by
$\epsilon \approx 1-2E_T/\sqrt{{\widehat s}}$, we get 
\begin{eqnarray}
\label{chayjet}{\frac{{d\sigma }}{{dE_T}}} \approx 
\bigl({\frac{{d\sigma }}{{dE_T}}}\bigr)_B
\bigl[1&-&{\frac{{4\alpha _s}}{{3\pi }}}\{3+4\ln 2(1-{\frac{{2E_T}
}{{\sqrt{{\hat s}}}}})\}\ln R \nonumber \\
&+&\cdots \bigr],
\end{eqnarray}
\noindent where $(d\sigma /dE_T)_B$ is the Born cross section in
Eq.\ (\ref{bornet}) and the dots represent
the terms irrelevant to the coefficient of $\ln R$. Note that this
only holds for small $R$ and for $E_T \approx \sqrt{{\widehat s}}/2$.
However, as we can see in Fig.\ \ref{figure2}, the logarithmic
dependence on $R$ is sustained for large $R$. From Eq.\
(\ref{chayjet}), the point where the coefficient of $\ln R$ vanishes
is given by $\ln (1-x_T) = -3/4-\ln 2=-1.44$ or $x_T \approx 0.76$, 
$E_T \approx 19$ GeV. Our numerical result shows 
that $b$ vanishes for $E_T \approx 18.5$ GeV or 
$x_T \approx 0.74$, which
is in (surprisingly) good agreement with this approximation. 
Said another way, we can understand the sign of $b$ from the
following considerations.  The $\ln R$ term arises from the 
perturbative collinear singularity integrated
over the angular phase space of the third parton.  When this parton is
inside the jet cone, $R$ appears as the {\em upper} limit of the angular 
integral and the coefficient of $\ln R$ is positive.  For the configuration
with the third parton outside of the jet cone, $R$ appears as the
{\em lower}
limit of the integral and $\ln R$ has a negative coefficient.  For large
$E_T$ values the former situation dominates and $b>0$, while for 
small $E_T$ the latter configuration is more important and $b<0$.

The coefficient $c$ is, in some sense, a measure of the contribution
from parts of the matrix element where the extra parton is essentially 
uncorrelated with the jet direction. This contribution should vary simply 
as the area of the cone,{\it \ i.e.}, as $R^2$.  We find that $c$ in
$e^{+}e^{-}$ collisions is small in magnitude compared to $a$ and
also $b$ (and, in fact, is difficult to fit reliably with our numerical
methods resulting in the small amplitude fluctuations in 
Fig.\ \ref{figure3}).  The $E_T$ variation of $c$ is understood in 
essentially the same way as for $b$.  Now the integral over the angular
phase space for the third parton yields the 2-dimensional area $R^2$
instead of $\ln R$.  
Thus we expect the coefficients 
$b$ and $c$ to exhibit very similar behavior as functions of $E_T$.
They are both negative at small $E_T$, positive at large $E_T$
({\it i.e.}, $E_T \approx \sqrt{{\hat s}}/2$)
and change sign at an intermediate $E_T$ value.  Our results suggest that they 
change sign at approximately the same $x_T$ ($E_T$)
value of about 0.75 to 0.8 ($E_T \approx 19$ GeV for 
$\sqrt{{\hat s}} = 50$ GeV).   The vanishing of the $R$ dependence
at a specific $E_T$ value was already apparent in Fig.\ \ref{figure1}.
This structure appears to be
characteristic of  $e^+ e^-$, at least in low order perturbation theory, 
and it 
will be interesting to check 
for it in the experimental data.

In hadronic collisions, the dependence on $R$ can be approximated by
the same form as in Eq.\ (\ref{rdepen}), but with different values of
the coefficients. We expect the same form because the logarithmic
term, which represents the collinear divergence at this order, appears
regardless of what the beam particles are. This dependence is a
general feature of the jet cross section in perturbation theory. 
However, the $E_T$ behavior of the coefficients
in $p{\overline p}$ collisions is expected to be 
quite different due to the kinematic
differences between the $e^+e^-$ and the $p{\overline p}$ cases 
that we discussed earlier. We expect that the coefficient $b$ is always
positive and does not vary much as we vary $E_T$ in the 
$p{\overline p}$ case.  This feature arises from the fact 
the partonic center-of-mass energy 
$\sqrt{{\widehat s}}(=$$\sqrt{x_1x_2 s})$ is almost always
just slightly larger
than $2E_T$.  This result is, in turn, ensured by the fact that the parton 
distribution functions are sharply peaked at
small $x$. Therefore $b$ is determined by the behavior of the cross
section at $2E_T/\sqrt{{\widehat s}}\approx 1$. It is obvious from the 
$e^+e^-$ case that $b$ should be positive in the $p{\overline p}$
case.   The same argument also suggests that the coefficient $c$ 
will be positive for all $E_T$ in the $p{\overline p}$ case. 
Furthermore the magnitude of $c$ should be relatively
larger in the $p{\overline p}$ case due to the contributions
from the initial-state radiation that is present in this case.
Since the initial-state radiation is correlated with the beam direction
and not the direction of the jet, the distribution of the partons from 
the initial-state radiation is rather isotropic with respect to the jet 
direction. 
Therefore the contribution of these partons to
the cross section is proportional to the area of the jet cone, $R^2$.
In the experimental data from $p{\overline p}$ collisions
one expects a further contribution to $c$
from the essentially uncorrelated
underlying event.  It will be informative to characterize
data from both  $e^+e^-$ and $p{\overline p}$ collisions
in terms of the coefficients in Eq.\ (\ref{rdepen}). 

\begin{table}
\caption{Comparison of  values for coefficients $a$, $b$ and $c$
for both $e^+ e^-$ and $p\overline{p}$ collisions scaled by the relevant
Born cross section.   All calculations are for $\mu = E_T/2$ and 
$R_{sep}=2 R$.  The $e^+ e^-$ results are for 
$\protect\sqrt{{\widehat s}}=50$ GeV while the $p\overline{p}$
numbers are for $\protect\sqrt{{\widehat s}}=1800$ GeV.}
\begin{center}
\begin{tabular}{cdddd}
Process & $E_T$(GeV) & $a/(d\sigma/dE_T)_{B}$ & $b/(d\sigma/dE_T)_{B}$ & 
$c/(d\sigma/dE_T)_{B}$ \\  \hline
$e^+ e^-$ & 10 & 0.68 & -0.80 & -0.12 \\
$e^+ e^-$ & 15 & 0.67 & -0.20 & -0.023 \\
$e^+ e^-$ & 18 & 0.77 & -0.035 & 0.014 \\
$e^+ e^-$ & 20 & 0.86 & 0.056 & 0.017 \\
$e^+ e^-$ & 22 & 0.92 & 0.15  & 0.041 \\
$e^+ e^-$ & 23 & 1.01 & 0.22 & 0.030 \\
$p\overline{p}$  & 100 & 0.74 & 0.18 & 0.27 \\
\label{table1}
\end{tabular}
\end{center}
\end{table}   

\medskip

The expectation that the $p{\overline p}$ jet cross section will 
increase rapidly with $R$ is illustrated for one $E_T$ value in 
Fig.~1 of the last paper in Ref. \cite{sellis}.  In the same paper 
an analysis in terms of Eq.\ (\ref{rdepen}) was carried out at 
$E_T$ = 100 GeV (see Table I of that paper).   To compare to 
the current study of theoretical cross sections, also with $\mu = 
E_T/2$ (and $R_{sep} = 2 R$), it is essential to scale out the 
overall differences between $e^+ e^-$ 
and $p{\overline p}$ by dividing out the Born cross section in 
each case.  The resulting scaled coefficients for a sampling of 
$E_T$ values are displayed in Table\ \ref{table1}.  As expected, the 
$p{\overline p}$ coefficients are all positive, of order 1 and 
most closely resemble the $e^+ e^-$ numbers for large values 
of $E_T$.  Note, in particular, the relatively good agreement of 
the $a$ and $b$ coefficients in the two processes for the largest 
$e^+ e^-$ $E_T$ values.  The relatively larger $a$ coefficient 
in the $e^+ e^-$ case is suggestive of narrower (and thus 
relatively $R$ independent) jets in 
$e^+ e^-$ collisions as will be discussed below.  Note also that the $c$ 
coefficient is larger in the $p\overline{p}$ case by approximately an order of 
magnitude as expected due to the presence of uncorrelated initial-state 
radiation.  Finally recall from Ref. \cite{sellis} that these 
theoretical results for 
$p\overline{p}$
are in approximate agreement with the experimental data.  A more detailed 
theoretical comparison, including the effects of  $R_{sep}$ will be presented 
elsewhere\cite{chayellis}.  Experimental studies over a broad $E_T$ range 
would also be useful.

\section{Jet Structure}

Now let us turn to the interesting question of the internal structure of 
jets.  In 
our idealized picture of the scattering process, contributions to the 
energy in 
the jet cone arise from final state radiation and, in the case of 
$p{\overline p}$ collisions, also from initial-state radiation and the 
underlying 
event.  Since the latter two contributions are uncorrelated with the jet 
direction, we naively expect jets in $p{\overline p}$ collisions to 
exhibit a less 
correlated, broader internal energy distribution compared to $e^+ e^-$
jets.  At the same time,
in both hadronic and $e^+e^-$ collisions, there is some energy that
falls outside the cone that is correlated to the hard scattering. The
higher-order calculation to first nontrivial order in $\alpha_s$
correctly includes the effect of the correlated energy that falls
outside the cone. In the hadronic case, it also accounts for the fact
that a large fraction of the energy far away from the jet is in fact
correlated, corresponding to soft but correlated perturbative
Bremsstrahlung in the hard event\cite{webber}. 

\begin{figure}
\vskip -0.9in
\centerline
     {  \epsfbox{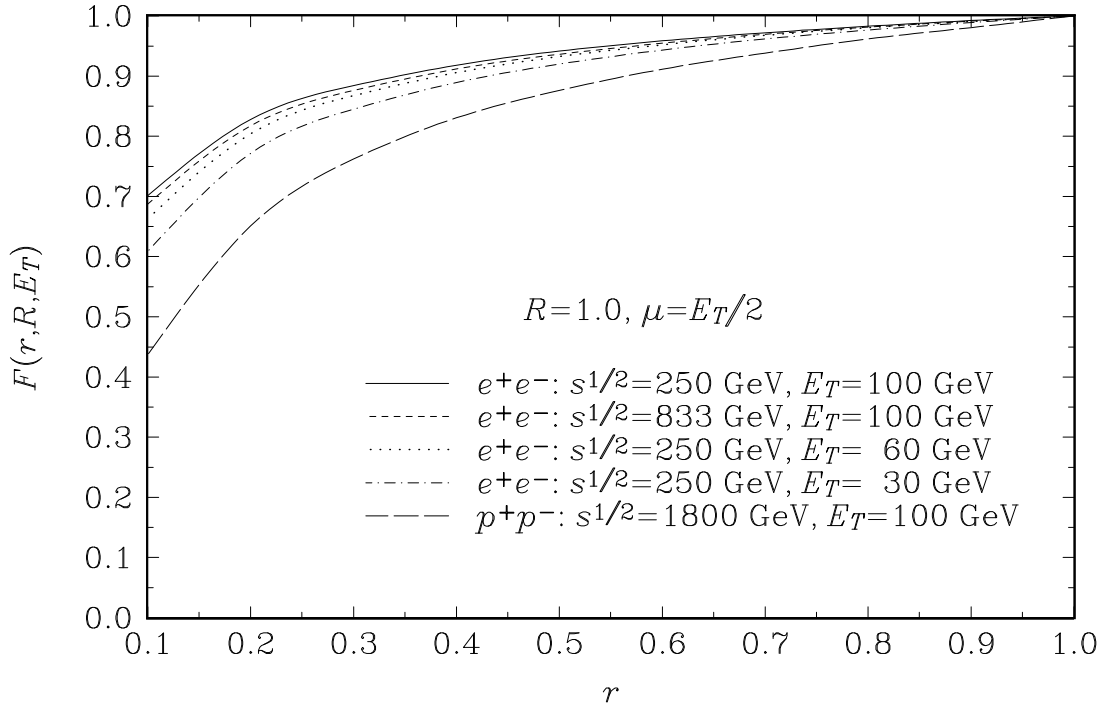}}
\vskip -0.25in
\caption{Fractional transverse energy distribution 
$F(r,R,E_T)$ for $r\leq 1.0$,
$R=1.0$ for various processes and values 
of $\protect\sqrt{{\widehat s}}$
and $E_T$.}
\label{figure4}
\end{figure}

It should be possible to confirm the validity of these ideas by
studying in detail the structure within jets in both the theoretical
results and in the experimental data. An example quantity is the
transverse energy distribution within a jet. Defining a jet sample
by a cone radius $R$ and a total jet transverse energy $E_T$, we
consider the fraction, $F(r, R, E_T)$ of $E_T$ that falls inside an
inner cone defined by the radius $r$. This quantity is constrained at the 
boundaries to be $F(0,R,E_T)=0$ and $F(R,R,E_T) = 1$.
The theoretical results of the
fraction $F(r,R, E_T)$ are illustrated in Fig.\ \ref{figure4}
corresponding to $e^+e^-$ jet samples with a few different values
of ${\sqrt{\widehat s}}$ and $E_T$, and a hadronic jet sample with
${\sqrt s} =$ 1800 GeV, $E_T=$ 100 GeV. (The last values are intended to be 
physically relevant to Tevatron data while the $e^+e^-$ 
values are chosen for easy comparison to the hadronic results.)  In both jet
samples, the jet cone is fixed at $R=1.0$. Most of the energy falls in
the small inner cone due to the collinear logarithmic contribution. If
we use the Born terms alone, all the energy will be at $r=0$ and 
$F(r>0,R, E_T)=1$ everywhere else inside the jet cone.  In contrast,
the calculation at order $\alpha_s$ exhibits a
nontrivial distribution, though there is still a (double) logarithmic
singularity for $r\rightarrow 0$. By comparing these two samples of jets
calculated from the theory, we can conclude that jets in
$e^+e^-$ collisions are narrower than jets in hadronic collisions.
That is, the fact that $F_{e^+ e^-} > F_{p \overline{p}}$ 
at all $r$ (for $0<r<R$) means that
the energy distribution inside of the jets in $e^+e^-$ collisions
is more concentrated near the center.  Note that this theoretical 
analysis does 
not include the further broadening that is expected to arise from the
underlying event contribution.

Fig.\ \ref{figure4} also illustrates that the jet shape in the $e^+ e^-$
jet sample varies systematically with $E_T$ and 
$\sqrt{{\widehat s}}$, becoming narrower with increasing $E_T$ at 
fixed $\sqrt{{\widehat s}}$ (compare $E_T$ = 30, 60 and 100 GeV at
$\sqrt{{\widehat s}}$ = 250 GeV) and broader with increasing 
$\sqrt{{\widehat s}}$ at fixed $E_T$ (compare 
$\sqrt{{\widehat s}}$ = 250 and 833 GeV at $E_T$ = 100 GeV).
But in the entire kinematic 
range, the jet from the $e^+e^-$ collision is consistently narrower 
than the hadronic jet.

To understand these results in more detail it is helpful 
to consider the quantity
$1-F(r, R, E_T)$, the distance {\it down} from the upper axis, 
which is the fraction of the transverse 
energy outside the smaller cone of radius $r$ and inside the jet size
$R$.  At the order we calculate the jet cross section this quantity is 
proportional to $\alpha_s(\mu)$ evaluated at $\mu = E_T/2$ and is
thus a decreasing function of $E_T$. This is directly illustrated by 
comparing the $e^+e^-$ jet sample with 
${\sqrt {\widehat s}} =$ 833 GeV and $E_T= 100$ GeV to the
sample with ${\sqrt {\widehat s}} =$ 250 GeV and $E_T=
$ 30 GeV.  In each sample the ratio $E_T/\sqrt {\widehat s}=0.12 $
is the same and the only
difference is the value of $\alpha_s (E_T/2)$. Since 
$1-F(r, R, E_T)$ is
proportional to $\alpha_s (E_T/2)$, the 20\% difference in
magnitudes of this quantity for the two samples arises from
the ratio of 
$\alpha_s (50 \ {\rm GeV})$ to $\alpha_s (15 \ {\rm GeV})$.

When we compare $e^+e^-$ jets to $p{\overline p}$ jets in 
Fig.\ \ref{figure4}, we find that the 
quantity $1-F(0.5,1.0, 100\ {\rm GeV})$ is about  2
times smaller for the former type of collision.  Again this illustrates
that 
the transverse energy is more concentrated near the center for a jet from
an $e^+e^-$ collision than for a jet from a hadronic collision. 
In these theoretical calculations this feature arises from two primary 
differences in the two kinds of collisions.  First, the 
initial-state radiation from
the beam hadrons in the hadronic case is absent in the $e^+e^-$
case. Though the partons produced in the initial-state radiation are 
uncorrelated with the direction of the final jets, they can come into
the jet cone and contribute to a broad jet energy distribution.   The 
second point that
contributes to the narrower jets in the $e^+e^-$ sample is that,
at order $\alpha_s$, the  $e^+e^-$ sample is dominated by naturally
narrower ``quark jets'', consisting of a quark plus a gluon.  By 
comparison the hadronic sample also includes a large component
of broader ``gluon jets'', consisting of two gluons (or a quark-antiquark
pair in the color octet state).

\begin{figure}
\vskip -1.05in
\centerline
    { \epsfbox{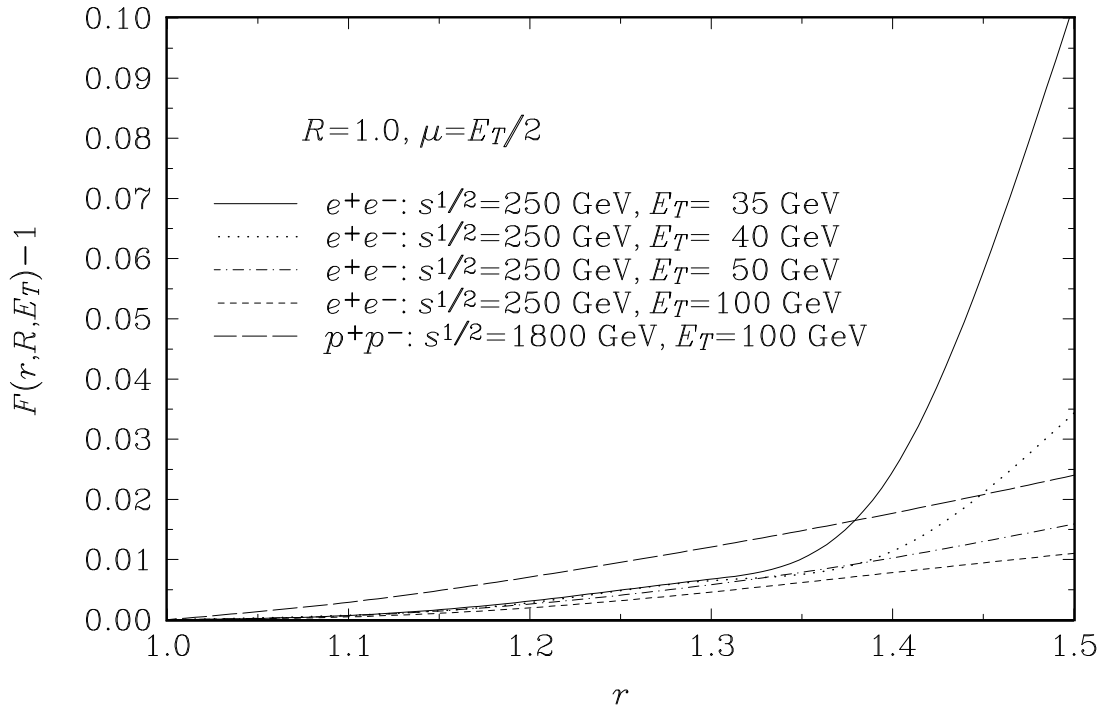} }
\vskip -0.25in
\caption{Fractional transverse energy distribution 
$F(r,R,E_T)-1.0$ for $r\geq 1.0$,
$R=1.0$ for various processes and values of $E_T$. }
\label{figure5}
\end{figure}

An interesting feature of the event structure correlated with jets
shows up when we consider the region $r>R$ as illustrated in Fig.\ 
\ref{figure5}.  In this region the quantity $F(r, R, E_T)-1$ measures how
much correlated
energy falls outside the jet.  Recall that $F(r,R,E_T)$ is the $E_T$
inside a cone of radius $r$ normalized to the fixed jet $E_T$ in the cone
of radius $R$. Thus the fractional $E_T$ {\em outside} of the jet
is given by $F(r,R,E_T)-F(R,R,E_T)=F(r,R,E_T)-1$. 
The correlated energy outside the jet cone comes  
from the partons produced after the hard scattering
for both the $e^+e^-$ and the $p{\overline p}$ cases, and
from those produced in the initial-state radiation before the hard
scattering for the $p{\overline p}$ case. The jet algorithm does not
distinguish where the partons come from.  It simply tries to identify
clusters of the partons that satisfy the criteria to form a jet. Based on the 
arguments used earlier, we might naively
expect that the energy outside of the cone in the hadronic case is larger
than that in the $e^+e^-$ case, since the hadronic collision
contains also the essentially isotropic initial-state radiation 
contribution. However, due to the kinematic differences in the two
situations, this simple expectation is not always realized.
The behavior for the $e^+e^-$ case depends strongly on the jet energy $E_T$. 
When $E_T$ is small compared to the fixed value of $\sqrt{{\widehat s}}$, 
$F(r,R,E_T)$ outside of the jet cone is necessarily large due to 
energy-momentum conservation.

As indicated in Fig.\ \ref{figure5}, $F(r,R,E_T)-1$ for the $e^+e^-$
case for $r>R$ with $\sqrt{{\widehat s}}=$ 250 GeV, remains small with 
small slope for $E_T \geq 50$ GeV but becomes a rapidly increasing
function of $r$ for $E_T < 40$ GeV.  The slope with $r$ systematically
increases with decreasing $E_T$.
It turns out that the behavior outside the jet for the hadronic jet sample 
does not 
vary much as we vary $E_T$ and the magnitude remains small (as indicated 
for the single $E_T$ value shown). This
relative independence of $E_T$ arises from a feature already discussed.
Since parton distribution functions are peaked at small values of
$x$,  $\sqrt {\hat s}$ is constrained to remain near its minimum value,
$2 E_T$ and the invariant mass squared of the final state resulting from the 
hard scattering stays small. If a parton outside the jet 
has appreciable energy, 
this would require a large invariant mass for the system composed
of that parton and the partons inside the jet. Therefore, if there is
a parton outside of the jet, the energy of that parton tends to be
small to make the combined invariant mass small. As an
example, the hadronic case with $\sqrt{s}=$ 1800 GeV, $E_T=$ 100 GeV
is shown in Fig.\ \ref{figure5}. 

Note that the transverse energy
deposited just outside the jet cone is very small in all cases. This feature
is due to the jet algorithm itself. The jet algorithm tries to include
as many partons as possible as long as the criteria for jet
formation, Eq.\ (\ref{bothcone}), are satisfied.  During the jet finding
process, the jet cone will tend to adjust its position in order to ``pull''
partons just outside of the cone into the cone.  In this way the $E_T$
inside the cone is increased.  This effect will play no role only when
the parton just outside the cone has vanishingly small $E_T$.
Therefore the transverse energy just outside the jet cone is 
very small in all cases.  
Of course, this feature is likely to be reduced by the largely uncorrelated, 
uniformly distributed contributions of the underlying event in the 
$p\overline{p}$ case and parton fragmentation effects in both 
$p\overline{p}$ and $e^+ e^-$ collisions.

\section{Conclusion}

The analysis of the inclusive jet cross section in $e^+e^-$ collisions
and in $p{\bar p}$ collisions, including the next-order contributions,
has achieved a high level of sophistication.  It is important that we further 
improve our understanding of physical processes involving jets so that we can 
use the jets as event tags to search for new physics.  
An essential issue is the 
systematic uncertainty that arises from the role played by the 
underlying event 
in processes involving hadrons in the initial-state.  To study this issue an 
important tool is the comparison of jets from different types of events where 
the underlying event is different. However, to study systematic effects of the 
jets in more detail, it is 
necessary to compare jet samples with the {\em same} jet definition
from the full range of possible processes, $p{\overline p}$
collisions, $e^+e^-$ annihilations and $ep$ scattering. 

Here we have theoretically
evaluated the jet cross section in both $e^+ e^-$ and $p\overline{p}$ 
collisions 
employing the same cone style 
 jet algorithm and have discussed the internal structure of the jets. 
By comparing theoretical jet cross sections from hadron collisions 
and $e^+e^-$ annihilations, we note differences in the global
dependence of the jet cross sections on the jet size $R$ and in the
distribution of the transverse energy inside the jets.  In particular, 
the jet cross 
section is expected to be a monotonically increasing function of $R$ for 
hadronic collisions while in $e^+e^-$ collisions the cross section is 
expected to 
increase with $R$ at large $x_T$ and decrease with $R$ at small $x_T$.  
However, in both cases we expect the $R$ dependence to exhibit the simple 
structure shown in Eq.\ (\ref{rdepen}), only the specific values of the 
coefficients will differ.
Perturbative QCD theory also predicts that the jets in $e^+e^-$ collisions are 
narrower than those produced in hadronic
collisions in the sense that the distribution of the transverse energy
inside the jet is more concentrated near the center in $e^+e^-$
collisions.  This difference has already been observed in a careful comparison
of $e^+ e^-$ and $p\overline{p}$ jet data performed by the OPAL 
Collaboration\cite{OPAL}.
In the current theoretical analysis these differences in jet cross sections 
and jet 
structures arise from the different kinematics discussed in Section 2, the 
presence of the
initial-state radiation in $p{\overline p}$ collisions, and the different 
mix of 
jet type (gluon versus quark) in the different processes.  A more thorough 
comparison of theory with data will be presented elsewhere\cite{chayellis}.

The calculation of the theoretical jet cross section
for $ep$ scattering is also in progress and it will be interesting to
compare all the jet cross sections from all of the available sources
of jets with the same jet definition. The case of $ep$ scattering is 
expected to 
exhibit characteristics in between the $e^+e^-$ and the $p{\overline p}$
cases since there is only a single hadron to serve as the source of the 
initial-state radiation and the underlying event.  However, as with the 
$e^+ e^-$ case, we have to be
careful about the kinematics.  For example, we have to identify the
energetic outgoing electrons to ensure that only photons and neutral 
weak $Z$ bosons are exchanged between the electron and the proton.
We hope that the comparison among these different jet samples will deepen
our understanding of the interplay of jet definitions with 
initial-state radiation and the underlying event 
as well as enhance our understanding 
of how different kinematic situations affect the jet cross sections.

\section*{Acknowledgments}

This work was supported by the U.S. Department of Energy under Grant
DE-FG06-91ER40614. One of the authors (J.C.) was supported in part by
KOSEF Grant 941-0200-02202, Ministry of Education BSRI 94-2408 and the 
Korea Science and Engineering Foundation through the SRC program of 
SNU-CTP.  One of us (S.D.E.) thanks J.W. Gary for several helpful 
discussions.


\begin{references}

\bibitem{ppbar}  See, for example, CDF Collaboration, F.~Abe {\it et al.},
Phys.~Rev.~Lett. {\bf 68}, 1109 (1992); {\it ibid.} {\bf 70}, 713 (1993); {\it 
ibid.} {\bf 70}, 1376 (1993).

\bibitem{ep}  See, for example, ZEUS Collaboration, Phys.~Lett.~B {\bf 306}, 
158 (1993);  Z.~Phys. C {\bf 59}, 231 (1993). 

\bibitem{ee}  See, for example, ``QCD Studies at LEP'', T.~Hebbeker,
{\it Proceedings of the Joint International Lepton-Photon Symposium \& 
Europhysics Conference On High Energy Physics}, Vol. 2, 73, edited by
S.~Hegarty et al., Geneva, 1991.

\bibitem{sellis}  S.D.~Ellis, Z.~Kunszt and D.E.~Soper, 
Phys.~\-Rev.~\-Lett.~{\bf 62}, 726 (1989);  Phys.~Rev.~D {\bf 40},
2188 (1989); Phys.~Rev.~Lett. {\bf 64}, 2121 (1990); {\it ibid.} {\bf 69}, 
3615 (1992).

\bibitem{greco} F. Aversa,   P. Chiappetta, M. Greco and J. Ph. Guillet,
 Phys.~Lett.~B {\bf  210}, 225 (1988); {\it ibid.} {\bf 211}, 465 (1988);
  Nucl. Phys.~B {\bf 327},105 (1989);
  Z.~Phys.~C {\bf46}, 235 (1990); {\it ibid.} {\bf 49}, 459 (1990);
  Phys.~Rev.~Lett. {\bf 65}, 401 (1990).

\bibitem{kellis}  R.K.~Ellis and J.C.~Sexton, Nucl.~Phys.~B {\bf
269}, 445 (1986).

\bibitem{snowmass}  See, for example, ``Towards a Standardization of
Jet Definitions'', J.~E.~Huth, {\it et al.}, {\it Proceedings of the 1990
DPF Summer Study on High Energy Physics}, Snowmass, 1990.

\bibitem{chayellis} J.~Chay, S.D.~Ellis and J.~Park, in preparation.

\bibitem{jetal}  For a compilation of jet algorithms, see B.~Flaugher
and K.~Meier, {\it Proceedings of the 1990 DPF Summer Study on High
Energy Physics}, Snowmass, 1990.

\bibitem{brown} L.S.~Brown and S.~Li, Phys.~Rev.~D {\bf 26}, 570
(1982).  

\bibitem{stwein}  G.~Sterman and S.~Weinberg,  Phys.~Rev.~Lett.
{\bf 39}, 1436 (1977).

\bibitem{webber}  G.~Marchesini and B.R.~Webber, Phys.~Rev. {\bf 38},
3419 (1988). 

\bibitem{OPAL} OPAL Collaboration, R. Akers, {\it et al.}, Z.~Phys.~C {\bf 
63}, 197 (1994).

\end{references}
\end{document}